\documentclass[10pt, journal , letterpaper]{IEEEtran}
\usepackage{amsmath,amssymb,amsfonts,amsthm,romannum}
\usepackage{algorithm,algorithmic}
\usepackage{graphicx,xcolor}
\usepackage[caption=false]{subfig}
\def\ie{{\it i.e.,\ \/}}

\DeclareMathOperator{\tr}{tr}
\DeclareMathOperator{\blkdiag}{blkdiag}
\theoremstyle{definition}
\newtheorem{theorem}{Theorem}

\newtheorem{lemma}{Lemma}

\makeatletter
\newcommand{\blue}{\textcolor{blue}}

\makeatother
\allowdisplaybreaks
\begin{document}

\pagenumbering{arabic}
\title
{\begin{Huge}Online Distributed Coordinated Precoding for Virtualized MIMO Networks with Delayed CSI\end{Huge}}
\author{
        Juncheng Wang, \IEEEmembership{Student Member, IEEE},
        Ben Liang, \IEEEmembership{Fellow, IEEE},
        Min Dong, \IEEEmembership{Senior Member, IEEE},\\       
        Gary Boudreau, \IEEEmembership{Senior Member, IEEE},
        and Hatem Abou-zeid, \IEEEmembership{Member, IEEE}\vspace{-5mm}
\thanks{J.~Wang and B.~Liang are with the University of Toronto (e-mail:
\{jcwang, liang\}@ece.utoronto.ca). M.~Dong is with the Ontario Tech University (e-mail: min.dong@ontariotechu.ca). G. Boudreau is with Ericsson Canada (email: gary.boudreau@ericsson.com). H. Abou-zeid was with Ericsson Canada and is now with the University of Calgary (email: hatem.abouzeid@ucalgary.ca).}
\thanks{ This work was supported in part by Ericsson Canada and by the Natural Sciences and Engineering Research Council of Canada.}
}
\maketitle

\begin{abstract}

We consider online wireless network virtualization (WNV) in a multi-cell multiple-input multiple output (MIMO) system with delayed feedback of channel state information (CSI). Multiple service providers (SPs) simultaneously share the base station resources of an infrastructure provider (InP). We aim at minimizing the accumulated precoding deviation of the InP's actual precoder from the SPs' virtualization demands via managing both inter-SP and inter-cell interference, subject to both long-term and short-term per-cell transmit power constraints. We develop an online coordinated precoding solution and show that it provides provable performance bounds. Our precoding solution is fully distributed at each cell, based only on delayed local CSI. Furthermore, it has a closed-form expression with low computational complexity. Finally, simulation results demonstrate the substantial performance gain of our precoding solution over the current best alternative.

\end{abstract}

\section{Introduction}
\label{Sec:1}

Multiple-input and multiple-output (MIMO) and wireless network virtualization (WNV) are widely recognized as two key technologies to meet the ever-increasing service demand in cellular networks. MIMO precoding enables a base station (BS) to serve multiple users simultaneously. Meanwhile, WNV allows multiple service providers (SPs) to share the BS resources of an infrastructure provider (InP), independent of the underlying physical infrastructure. In WNV, the InP virtualizes the physical resources into virtual slices, while each SP leases some of these virtual slices to provide services to its own end users. Different from wired network virtualization, it is challenging to guarantee service isolation among these SPs, due to the broadcast and fading nature of wireless channels~\cite{WNV-M.Richart'16}.

Early works on MIMO WNV promote the allocation of orthogonal subchannels or exclusive subsets of antennas among the SPs for service isolation \cite{WNV-V.Jumba'15}\nocite{WNV'K.Zhu'16}-\cite{WNV-Y.Liu'18}. Such physical separation is directly inherited from wired network virtualization and does not fully utilize MIMO antennas for spatial multiplexing.  In contrast, the \textit{spatial} virtualization approach in \cite{WNV-M.Soltanizadeh'18} isolates the SPs via MIMO precoding at the InP, allowing simultaneous sharing of all spectrum and antennas among the SPs.

In practical wireless systems, \textit{long-term} transmit power is an important measure of energy efficiency \cite{MIMO-H.Q.Ngo'2013}. Under a time-averaged
transmit power limit, MIMO precoding design for WNV becomes an \textit{online} optimization problem, dependent on the underlying time-varying channels. Therefore, recent works have extended \cite{WNV-M.Soltanizadeh'18} to the online setting with instantaneous and one-slot delayed channel state information (CSI) in \cite{WNV-J.Wang'INFOCOM20} and \cite{WNV-J.Wang'SPAWC20}, respectively. However, in a MIMO system with many transmit antennas, the CSI can be severely delayed for \textit{multiple} transmission frames, due to the need for channel estimation, quantization, and feedback. Furthermore, the above mentioned works on WNV all focus on single-cell MIMO systems.

In non-virtualized networks, multi-cell \textit{coordinated} transmission
is known to substantially outperform non-coordinated transmission, as a result
of efficient inter-cell interference mitigation \cite{COOD-Dahrouj'2010}.
Compared with multi-cell cooperative transmission \cite{COOP-D.Gesbert'2010},
coordinated precoding does not require data sharing or stringent symbol-level
synchronization among BSs. Most of works on multi-cell coordinated
precoding consider the problem as deterministic per-slot optimization \cite{COOD-Dahrouj'2010},~\cite{S.He:COMP'14}.
Only a few works adopt online approaches for stochastic coordinated precoding
design \cite{R.Kim:OCOMP'19},~\cite{J.Ge:OCOMP'20}. For virtualized networks,
a per-slot coordinated precoding design for multi-cell MIMO WNV with perfect
CSI has been proposed in \cite{WNV-J.Wang'TWC2021} under short-term transmit
power constraints. Other design approaches, such as resource allocation
or pricing, have also been considered for multi-cell WNV \cite{MC-K.Ahsan'18},~\cite{MC-T.Duy'18}.
These works focus on per-slot design and do not consider MIMO in WNV.

In this work, we consider an online coordinated precoding design for WNV in a multi-cell MIMO system, with CSI feedbacks that are possibly delayed for multiple time slots. In each cell, each SP designs its virtual precoder for its own users, without the knowledge of either inter-SP or inter-cell interference. The InP designs the actual coordinated precoder to meet the SPs' virtual precoding demands over time while managing the interference among the SPs and cells, subject to both long-term and short-term transmit power constraints at each cell. We note that due to the long-term transmit power constraints, the coordinated precoders are correlated over time, and the resulting online problem is particularly challenging to solve due to CSI feedback delays.

The main contributions of this letter are summarized below:

\noindent$\bullet$ We formulate the coordinated multi-cell MIMO WNV problem as a constrained online convex optimization (OCO) problem with multi-slot feedback delay. At each time slot, the InP designs a coordinated precoder to meet the SPs' virtualization demands, under both instantaneous and time-averaged transmit power limits. 

\noindent$\bullet$ We develop an online  coordinated precoding solution for this problem, which is inspired by our recent work on general delay-tolerant OCO \cite{COCO-J.Wang'21}. We show that it has provable performance bounds. Unlike the solution in \cite{COCO-J.Wang'21}, the proposed precoding solution is fully distributed without any CSI exchange among BSs. Moreover, the precoder solution at each time slot is given in a closed-form expression, which implies low computational complexity for implementation.

\noindent$\bullet$ Simulation results of our precoding solution under typical urban micro-cell Long-Term Evolution (LTE) network settings demonstrate substantial performance advantage over the current best alternative.

\section{System Model and Problem Formulation}

\label{Sec:2}

\subsection{System Model}
\label{Sec:2A}

We consider an InP that  performs  downlink WNV in a MIMO network consisting of $C$ cells. In each cell $c$, the InP owns a BS equipped with $N_c$ antennas. Thus, there is a total of $N=\sum_{c=1}^CN_c$ antennas in the network. The InP serves $M$ SPs. Each SP $m$ has $K_c^m$ subscribing users in cell $c$. There is a total of $K_c=\sum_{m=1}^MK_c^m$ users in cell $c$ and $K=\sum_{c=1}^CK_c$ users in the network. We consider a time-slotted system with time slot indexed by $t$. Let $\mathbf{H}_t^{lc,m}\in\mathbb{C}^{K_l^m\times{N}_c}$ be the channel state between the $K_l^m$ users of SP $m$ in cell $l$ and the BS $c$ at time~$t$.

\subsubsection{Multi-cell WNV} 
\begin{figure}[t]
\vspace{-0mm}
\centering
\includegraphics[width=.8\linewidth,trim= 230 140 200 140,clip]{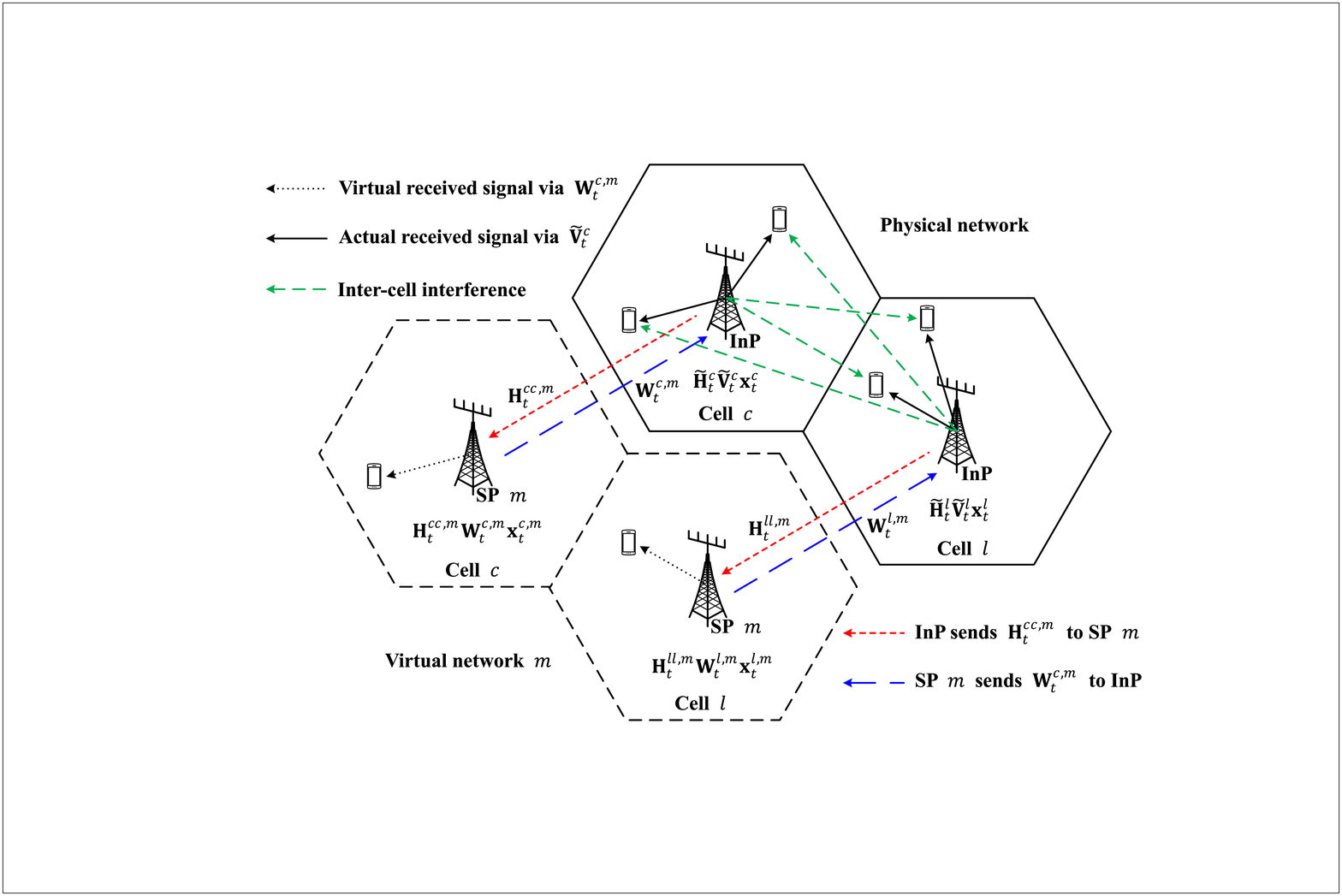}
\vspace{-4mm}
\caption{An illustration of downlink coordinated multi-cell MIMO WNV.}
\label{fig:1}
\vspace{-2mm}
\end{figure}

We first illustrate our multi-cell MIMO WNV framework with coordinated precoding in the idealized scenario without CSI feedback delay, as shown in Fig.~\ref{fig:1}.

At each time $t$, in each cell $c$, the InP shares the corresponding CSI $\mathbf{H}_t^{cc,m}\in\mathbb{C}^{K_c^m\times{N}_c}$ with each SP $m$, and allocates \textit{virtual} transmit power $P_c^m$ to the SP. Then, each SP~$m$ designs its own \textit{virtual} precoder $\mathbf{W}_t^{c,m}\in\mathbb{C}^{N_c\times{K}_c^m}$, under the virtual transmit power limit $\Vert\mathbf{W}_t^{c,m}\Vert_{F}^2\le{P}_c^m$, where $\Vert\cdot\Vert_F$ denotes the Frobenius norm. Note that each SP $m$ designs $\mathbf{W}_t^{c,m}$ based on the service needs of its own users, without the knowledge of the other SP's users in the cell or the users in other cells. Each SP $m$ then sends $\mathbf{W}_t^{c,m}$ to the InP as its service demand in cell $c$. 

With the virtual precoders $\{\mathbf{W}_t^{c,m}\}$ demanded by the SPs, the \textit{virtual} received (noiseless) signal vector at the $K_c^m$ users of SP $m$ in cell $c$ is given by
\begin{align*}
        \tilde{\mathbf{y}}_t^{c,m}=\mathbf{H}_t^{cc,m}\mathbf{W}_t^{c,m}\mathbf{x}_t^{c,m},\quad\forall{m},\forall{c}
\end{align*}
where $\mathbf{x}_t^{c,m}$ is the transmitted signal vector. Let  $\tilde{\mathbf{y}}_t^c=[\hbox{$\tilde{\mathbf{y}}_t^{c,1}$}^H,\dots,\hbox{$\tilde{\mathbf{y}}_t^{c,M}$}^H]^H$ be the virtual received signal vector at the $K_c$ users in cell $c$. We have $\tilde{\mathbf{y}}_t^c=\mathbf{D}_t^c\mathbf{x}_t^c$, where $\mathbf{D}_t^c\triangleq\blkdiag\{\mathbf{H}_t^{cc,1}\mathbf{W}_t^{c,1},\dots,\mathbf{H}_t^{cc,M}\mathbf{W}_t^{c,M}\}\in\mathbb{C}^{K_c\times{K_c}}$ is the virtualization demand from cell $c$ and $\mathbf{x}_t^c=[\hbox{$\mathbf{x}_t^{c,1}$}^H,\dots,\hbox{$\mathbf{x}_t^{c,M}$}^H]^H$. Denote $\tilde{\mathbf{y}}_t=[\hbox{$\tilde{\mathbf{y}}_t^1$}^H,\dots,\hbox{$\tilde{\mathbf{y}}_t^C$}^H]^H$ as the virtual received signal vector at all $K$ users in the network. We have $\tilde{\mathbf{y}}_t=\mathbf{D}_t\mathbf{x}_t$, where $\mathbf{D}_t\triangleq\blkdiag\{\mathbf{D}_t^1,\dots,\mathbf{D}_t^C\}\in\mathbb{C}^{K\times{K}}$ and $\mathbf{x}_t=[\hbox{$\mathbf{x}_t^1$}^H,\dots,\hbox{$\mathbf{x}_t^C$}^H]^H$. The transmitted signals are assumed to be independent of each other with unit power, \ie $\mathbb{E}\{\mathbf{x}_t\mathbf{x}_t^H\}=\mathbf{I},\forall{t}$.

Let $\bar{\mathbf{H}}_t^{lc}=[\hbox{$\mathbf{H}_t^{lc,1}$}^H,\dots,\hbox{$\mathbf{H}_t^{lc,M}$}^H]^H\in\mathbb{C}^{K_l\times{N}_c}$
be the channel state between the $K_l$ users in cell $l$ and the BS $c$. In each cell $c$, based on local CSI $\tilde{\mathbf{H}}_t^c=[\hbox{$\bar{\mathbf{H}}_t^{1c}$}^H,\dots,\hbox{$\bar{\mathbf{H}}_t^{Cc}$}^H]^H\in\mathbb{C}^{K\times{N}_c}$, the InP designs the \textit{actual} precoder $\tilde{\mathbf{V}}_t^c=[\mathbf{V}_t^{c,1},\dots,\mathbf{V}_t^{c,M}]\in\mathbb{C}^{N_c\times{K}_c}$ to serve the $K_c$ users, where $\mathbf{V}_t^{c,m}\in\mathbb{C}^{N_c\times{K}_c^m}$ is the precoder designed for SP $m$. The \textit{actual} received (noiseless) signal vector at the $K_c^m$ users of SP $m$ in cell $c$ is given by
\begin{align*}
        \mathbf{y}_t^{c,m}&=\mathbf{H}_t^{cc,m}\mathbf{V}_t^{c,m}\mathbf{x}_t^{c,m}+\sum_{i=1,i\neq{m}}^M\mathbf{H}_t^{cc,m}\mathbf{V}_t^{c,i}\mathbf{x}_t^{c,i}\notag\\
        &\quad+\sum_{l=1,l\neq{c}}^C\sum_{j=1}^M\mathbf{H}_t^{cl,j}\mathbf{V}_t^{l,j}\mathbf{x}_t^{l,j},\quad\forall{m},\forall{c}
\end{align*}
where the second term is the inter-SP interference caused by the other SP's users in cell $c$, and the third term is the inter-cell interference caused by the users in other cells. Let $\mathbf{y}_t=[\hbox{$\mathbf{y}_t^1$}^H,\dots,\hbox{$\mathbf{y}_t^C$}^H]^H$ be the actual received signal vector at all $K$ users in the network, where $\mathbf{y}_t^c=[\hbox{$\mathbf{y}_t^{c,1}$}^H,\dots,\hbox{$\mathbf{y}_t^{c,M}$}^H]^H$.
We have $\mathbf{y}_t=\mathbf{H}_t\mathbf{V}_t\mathbf{x}_t$, where $\mathbf{H}_t=[\tilde{\mathbf{H}}_t^1,\dots,\tilde{\mathbf{H}}_t^C]\in\mathbb{C}^{K\times{N}}$ is the global channel state and $\mathbf{V}_t=\blkdiag\{\tilde{\mathbf{V}}_t^1,\dots,\tilde{\mathbf{V}}_t^C\}\in\mathbb{C}^{N\times{K}}$ is the actual global precoder.

\subsubsection{Delayed CSI} In practical multi-cell MIMO networks, instantaneous CSI is usually unavailable to the InP. Typically, at each time $t$, the InP only has the $\tau$-slot delayed CSI $\tilde{\mathbf{H}}_{t-\tau}^c$ at each cell~$c$, where $\tau\ge1$ is the CSI feedback delay. Thus, each SP~$m$ only has the delayed CSI $\mathbf{H}_{t-\tau}^{cc,m}$ to design its own virtual precoder $\mathbf{W}_{t-\tau}^{c,m}$ at each cell $c$. As a result, the InP receives a delayed virtualization demand $\mathbf{D}_{t-\tau}^c$ from each cell~$c$. Using $\tilde{\mathbf{H}}_{t-\tau}^c$ and $\mathbf{D}_{t-\tau}^c$, the InP designs the actual precoder $\tilde{\mathbf{V}}_t^c$ for each cell~$c$.

\subsection{Problem Formulation}

The InP coordinates the cells to design the actual global precoder to meet the virtualization demand gathered from the SPs, while implicitly eliminating both inter-SP and inter-cell interference. The expected deviation of received signals at all $K$ users in the network, for the InP's precoder $\mathbf{V}_t$ and the SPs' virtualization demand $\mathbf{D}_t$ is $\mathbb{E}_{\mathbf{x}_t}\{\Vert\mathbf{y}_t-\tilde{\mathbf{y}}_t\Vert^2\}=\Vert\mathbf{H}_t\mathbf{V}_t-\mathbf{D}_t\Vert_F^2=\sum_{c=1}^C\Vert\tilde{\mathbf{H}}_t^c\tilde{\mathbf{V}}_{t}^c-\tilde{\mathbf{D}}_{t}^c\Vert_F^2$,
where $\tilde{\mathbf{D}}_t^c\triangleq[\mathbf{0},\dots,{\hbox{$\mathbf{D}_t^c$}}^H,\dots,\mathbf{0}]^H\in\mathbb{C}^{K\times{K}^c}$ and $\Vert\cdot\Vert$ is the Euclidean norm. We define the deviation of InP's precoder from the SPs' virtualization demand as
\begin{align}
        f_t(\mathbf{V}_t)\triangleq\Vert\mathbf{H}_t\mathbf{V}_t-\mathbf{D}_t\Vert_F^2,\quad\forall{t}
\end{align}
which is a convex loss function.

For a total of $T$ time slots, we assume the following long-term transmit power constraint at each BS $c$:
\begin{align}
        \frac{1}{T}\sum_{t=1}^T\Vert\tilde{\mathbf{V}}_t^c\Vert_F^2\le\bar{P}_c,\quad\forall{c}\label{eq:LTCc}
\end{align}
where  $\bar{P}_c$ is the average transmit power limit. We also consider short-term transmit power constraints, collectively represented by a convex feasible set $\mathcal{V}\triangleq\{\mathbf{V}_t:\Vert\tilde{\mathbf{V}}_t^c\Vert_F^2\le{P}_c^{\text{max}},\forall{c}\}$,
where $P_c^{\text{max}}$ is the maximum transmit power limit at BS $c$. 

The goal at the InP is to optimize the MIMO precoders to minimize the accumulated precoding deviation over time in the presence of delayed CSI, subject to both long-term and short-term transmit power constraints at each cell. The optimization problem is formulated as a constrained OCO problem as follows:
\begin{align}
        \textbf{P1}:\quad\min_{\{\mathbf{V}_t\in\mathcal{V}\}}\quad&\sum_{t=1}^Tf_t(\mathbf{V}_t)\notag\\
        \text{s.t.}\qquad&\sum_{t=1}^T\mathbf{g}(\mathbf{V}_t)\preceq\mathbf{0}\label{eq:LTC}
\end{align}
where $\mathbf{g}(\mathbf{V}_t)=[g^1(\tilde{\mathbf{V}}_t^1),\dots,g^C(\tilde{\mathbf{V}}_t^C)]^T$
with $g^c(\tilde{\mathbf{V}}_t^c)\triangleq\Vert\tilde{\mathbf{V}}_t^c\Vert_F^2-\bar{P}_c$. Note that constraints (\ref{eq:LTCc}) and (\ref{eq:LTC}) are equivalent.

In this work, without assuming knowledge on the channel distribution, we aim at developing an online coordinated precoding solution $\{\mathbf{V}_t\}$ to $\textbf{P1}$, based on the $\tau$-slot delayed CSI $\mathbf{H}_{t-\tau}$ and virtualization demand $\mathbf{D}_{t-\tau}$.

\section{Online Coordinated Multi-Cell MIMO WNV}
\label{Sec:3}

In this section, we present an online coordinated multi-cell MIMO WNV algorithm that is inspired by our general delay-tolerant OCO algorithm in \cite{COCO-J.Wang'21}. Note that the online algorithm in \cite{COCO-J.Wang'21} is centralized. In contrast, our online algorithm for solving $\textbf{P1}$ is \textit{fully distributed} at each cell without any CSI exchange among cells, and we further provide a closed-form solution to each per-slot coordinated precoding optimization problem.

\subsection{Fully Distributed Online Solution Framework}

We first introduce a virtual queue vector $\mathbf{Q}_t=[Q_t^1,\dots,Q_t^C]^T$ for the long-term transmit power constraints in (\ref{eq:LTC}), with the following dynamics:
\begin{align}
        Q_t^c=\max\{-\gamma{g}^c(\tilde{\mathbf{V}}_t^c),Q_{t-1}^c+\gamma{g}^c(\tilde{\mathbf{V}}_t^c)\},\quad\forall{c}\label{eq:VQ}
\end{align}
where $\gamma>0$ is a weighting factor on the constraint violation that affects how much the virtual queue changes over time. The virtual queue works like a Lagrangian multiplier vector for $\textbf{P1}$ or a backlog queue for the constraint violation. We then convert $\textbf{P1}$ to a per-slot problem at each slot $t>\tau$, subject to the short-term transmit power constraints only, given by
\begin{align*}
        \textbf{P2}:~\min_{\mathbf{V}\in\mathcal{V}}~&2\Re\{\tr\{[\nabla_{\mathbf{V}_{t-\tau}^*}{f}_{t-\tau}(\mathbf{V}_{t-\tau})]^H(\mathbf{V}-\mathbf{V}_{t-\tau})\}\}\notag\\
        &\quad+[\mathbf{Q}_{t-1}+\gamma\mathbf{g}(\mathbf{V}_{t-1})]^T[\gamma\mathbf{g}(\mathbf{V})]\notag\\
        &\quad+\alpha\Vert\mathbf{V}-\mathbf{V}_{t-\tau}\Vert_F^2+\eta\Vert\mathbf{V}-\mathbf{V}_{t-1}\Vert_F^2
\end{align*}
where $\alpha,\eta>0$ are two step-size parameters that control the weights on the two regularization terms, $\Re\{\cdot\}$ denotes the real part of the enclosed parameter, $\tr\{\mathbf{A}\}$ denotes the trace of matrix $\mathbf{A}$, and $\nabla_{\mathbf{V}_{t-\tau}^*}{f}_{t-\tau}(\mathbf{V}_{t-\tau})=\mathbf{H}_{t-\tau}^H(\mathbf{H}_{t-\tau}\mathbf{V}_{t-\tau}-\mathbf{D}_{t-\tau})$
is the partial derivative of $f_{t-\tau}(\mathbf{V}_{t-\tau})$ with respect
to (w.r.t.) the complex conjugate of $\mathbf{V}_{t-\tau}$.

Compared with the original $\textbf{P1}$, the long-term transmit power constraint (\ref{eq:LTC}) has been moved into the objective function in $\textbf{P2}$ as a penalization term. Note that \textbf{P2} uses double regularization $\alpha\Vert\mathbf{V}-\mathbf{V}_{t-\tau}\Vert_F^2$ and $\eta\Vert\mathbf{V}-\mathbf{V}_{t-1}\Vert_F^2$, which was first proposed in \cite{COCO-J.Wang'21}. The intuition behind the double regularization is that both $\mathbf{V}_{t-\tau}$ and $\mathbf{V}_{t-1}$ help  minimize the accumulated precoding deviation and the violation of long-term transmit power. Therefore, it is desirable to keep the new precoder $\mathbf{V}_t$ at time $t$ close to both $\mathbf{V}_{t-\tau}$ and $\mathbf{V}_{t-1}$. 

Note that $\mathbf{V}$ and $\nabla_{\mathbf{V}_{t-\tau}^*}{f}_{t-\tau}(\mathbf{V}_{t-\tau})$ in \textbf{P2} are block diagonal matrices, with the $c$-th block associated with the precoder for cell $c$. In addition, $\mathbf{g}(\mathbf{V})$ and $\mathcal{V}$ are separable among cells. Thus, $\textbf{P2}$ can be \textit{equivalently} decomposed into $C$ subproblems $\{\textbf{P3}^c\}$, each corresponding to the local precoder optimization problem for $\tilde{\mathbf{V}}_t^c$ at cell $c$, given by
\begin{align}
        \textbf{P3}^c:~\min_{\tilde{\mathbf{V}}^c}~&2\Re\{\tr\{[\nabla_{\tilde{\mathbf{V}}_{t-\tau}^{c*}}{f}_{t-\tau}(\mathbf{V}_{t-\tau})]^H(\tilde{\mathbf{V}}^c-\tilde{\mathbf{V}}_{t-\tau}^c)\}\}\notag\\
        &\quad+[Q_{t-1}^c+\gamma{g}^c(\tilde{\mathbf{V}}^c_{t-1})]\gamma{g}^c(\tilde{\mathbf{V}}^c)\notag\\
        &\quad+\alpha\Vert\tilde{\mathbf{V}}^c-\tilde{\mathbf{V}}_{t-\tau}^c\Vert_F^2+\eta\Vert\tilde{\mathbf{V}}^c-\tilde{\mathbf{V}}^c_{t-1}\Vert_F^2\notag\\
        \text{s.t.}~~&\Vert\tilde{\mathbf{V}}^c\Vert_F^2\le P_c^{\text{max}}\label{eq:STC}
\end{align}
where $\nabla_{\tilde{\mathbf{V}}_{t-\tau}^{c*}}f_{t-\tau}(\mathbf{V}_{t-\tau})=\tilde{\mathbf{H}}_{t-\tau}^{cH}(\tilde{\mathbf{H}}_{t-\tau}^c\tilde{\mathbf{V}}_{t-\tau}^c-\tilde{\mathbf{D}}_{t-\tau}^c)$. At each time $t>\tau$, based on the delayed local CSI $\tilde{\mathbf{H}}_{t-\tau}^c$ and virtualization demand $\tilde{\mathbf{D}}_{t-\tau}^c$, the InP obtains the current local precoder $\tilde{\mathbf{V}}_t^c$ by solving $\textbf{P3}^c$ for each cell $c$. Therefore, the per-slot coordinated precoder optimization problem $\textbf{P2}$ leads to a fully-distributed implementation at each cell, without any CSI exchange among cells.

\subsubsection*{Summary of our online solution framework} 1)~Initialize $\alpha,\eta,\gamma>0$, $\mathbf{V}_t\in\{\mathbf{V}:\mathbf{g}(\mathbf{V})=\mathbf{0}\}$
and $\mathbf{Q}_t=\mathbf{0}$, for any $t\le\tau$; 2)~At each time $t>\tau$, obtain $\tilde{\mathbf{H}}_{t-\tau}^c$ and $\tilde{\mathbf{D}}_{t-\tau}^c$ at each cell $c$, and update $\tilde{\mathbf{V}}_t^c$ by solving $\textbf{P3}^c$ via (\ref{eq:Vtc}) presented in Section~\ref{Sec:3B}; 3) Update the virtual queue $Q_t^c$ for each cell $c$ via (\ref{eq:VQ}). The choice of $\alpha,\eta,\gamma$ will be discussed in Section~\ref{Sec:3C}, when we derive the performance bounds for our online precoding solution.

\subsection{Online Precoding Solution to $\textbf{P3}^c$}
\label{Sec:3B}

Now we solve $\textbf{P3}^c$ to obtain the precoder $\tilde{\mathbf{V}}_t^c$ in each cell $c$. Note that $\textbf{P3}^c$ is a convex optimization problem with strong duality. We solve it by using the Karush-Kuhn-Tucker (KKT) conditions. The Lagrangian for $\textbf{P3}^c$ is
\begin{align*}
        L(\tilde{\mathbf{V}}^c,\lambda^c)&=2\Re\{\tr\{[\nabla_{\tilde{\mathbf{V}}_{t-\tau}^{c*}}f_{t-\tau}(\mathbf{V}_{t-\tau})]^H(\tilde{\mathbf{V}}^c-\tilde{\mathbf{V}}_{t-\tau}^c)\}\}\notag\\
        &~+[Q_{t-1}^c\!+\!\gamma{g}^c(\tilde{\mathbf{V}}_{t-1}^c)]\gamma{g}^c(\tilde{\mathbf{V}}^c)+\alpha\Vert\tilde{\mathbf{V}}^c\!-\!\tilde{\mathbf{V}}_{t-\tau}^c\Vert_F^2\notag\\
        &~+\eta\Vert\tilde{\mathbf{V}}^c-\tilde{\mathbf{V}}_{t-1}^c\Vert_F^2+\lambda^c(\Vert\tilde{\mathbf{V}}^c\Vert_F^2-P_c^{\text{max}})
\end{align*}
where $\lambda^c$ is the Lagrange multiplier associated with the short-term transmit power constraint (\ref{eq:STC}).


The KKT conditions for the globally optimal $(\tilde{\mathbf{V}}^{c\circ},\lambda^{c\circ})$
are $\Vert\tilde{\mathbf{V}}^{c\circ}\Vert_F^2- P_{\text{max}}^c\le0$, $\lambda^{c\circ}\ge0$, $\lambda^{c\circ}(\Vert\tilde{\mathbf{V}}^{c\circ}\Vert_F^2-P_c^{\text{max}})=0$, and  
\begin{align}
        \tilde{\mathbf{V}}^{c\circ}=\frac{\alpha\tilde{\mathbf{V}}_{t-\tau}^c+\eta\tilde{\mathbf{V}}_{t-1}^c-\tilde{\mathbf{H}}_{t-\tau}^{cH}(\tilde{\mathbf{H}}_{t-\tau}^c\tilde{\mathbf{V}}_{t-\tau}^c-\tilde{\mathbf{D}}_{t-\tau}^c)}{\gamma{Q}_{t-1}^c+\gamma^2{g}^c(\tilde{\mathbf{V}}_{t-1}^c)+\alpha+\eta+\lambda^{c\circ}},\!\label{eq:KKT}
\end{align}
which follows from setting the partial derivative $\nabla_{\tilde{\mathbf{V}}^{c*}}L(\tilde{\mathbf{V}}^c,\lambda^c)$
to $\mathbf{0}$. From the KKT conditions, and noting that $\lambda^{c\circ}$ can be seen as a power scaling factor for $\tilde{\mathbf{V}}^{c\circ}$ in (\ref{eq:KKT}), we have a closed-form solution for $\tilde{\mathbf{V}}_t^c$,
given by
\begin{align}
        \tilde{\mathbf{V}}_t^c=\left\{\begin{matrix}
        \mathbf{X}_t^c,&\text{if~}\Vert\mathbf{X}_t^c\Vert_F^2\leq P_c^{\text{max}}\\
        \frac{\sqrt{P_c^{\text{max}}}}{\Vert\mathbf{X}_t^c\Vert_F}\mathbf{X}_t^c,&\text{o.w.}
        \end{matrix}\right.\label{eq:Vtc}
\end{align}
where $\mathbf{X}_t^c=\frac{\alpha\tilde{\mathbf{V}}_{t-\tau}^c+\eta\tilde{\mathbf{V}}_{t-1}^c-\tilde{\mathbf{H}}_{t-\tau}^{cH}(\tilde{\mathbf{H}}_{t-\tau}^c\tilde{\mathbf{V}}_{t-\tau}^c-\tilde{\mathbf{D}}_{t-\tau}^c)}{\gamma{Q}_{t-1}^c+\gamma^2{g}^c(\tilde{\mathbf{V}}_{t-1}^c)+\alpha+\eta}$.

Note that the computational complexity of calculating $\tilde{\mathbf{V}}_t^c$ is dominated by matrix multiplication, and thus is in the order of $\mathcal{O}(KN_cK_c)$. Furthermore, per-antenna maximum transmit power limits can be incorporated in the short-term transmit power constraint (\ref{eq:STC}). In this case, $\textbf{P3}^c$ can be equivalently decomposed into $N_c$ subproblems, each with a closed-form solution similar to (\ref{eq:Vtc}).

\subsection{Performance Bounds}
\label{Sec:3C}

Due to the lack of instantaneous CSI, it is impossible for the InP to obtain an optimal precoding solution to the constrained OCO problem $\textbf{P1}$. A widely adopted performance measure in the OCO literature is the \textit{regret}, given by
\begin{align}
        \text{RE}(T)\triangleq\sum_{t=1}^T\left(f_t(\mathbf{V}_t)-f_t(\mathbf{V}^\star)\right)\label{eq:regdef}
\end{align}
where $\mathbf{V}^\star\triangleq\arg\min_{\mathbf{V}\in\mathcal{V}}\{\sum_{t=1}^Tf_t(\mathbf{V})|\mathbf{g}(\mathbf{V})\preceq\mathbf{0}\}$ is the standard offline fixed solution to \textbf{P1} assuming all the CSI over all time $\{\mathbf{H}_t\}$ is known apriori. Furthermore, to measure the accumulated violation of the long-term transmit power constraints, we define the \textit{constraint violation} as
\begin{align}
        \text{VO}^c(T)\triangleq\sum_{t=1}^Tg^c(\tilde{\mathbf{V}}_t^c),\quad\forall{c}.\label{eq:viodef}
\end{align}

We now provide performance bounds for our online precoding solution. For performance analysis, we assume the channel gain at any time $t$ is upper bounded by a constant $B>0$, given by $\Vert\mathbf{H}_t\Vert_F\le{B},\forall{t}$. The following lemma shows that $\textbf{P1}$ satisfies several general assumptions in the OCO literature: 1) The gradient of the convex loss function $\nabla_{\mathbf{V}^*} f_t(\mathbf{V})$ is bounded; 2) The long-term constraint function $\mathbf{g}(\mathbf{V})$ is Lipschitz continuous; 3) The impact of $\mathbf{g}(\mathbf{V})$ is bounded; 4) The impact of the convex set $\mathcal{V}$ is bounded; 5)~There exists an interior point $\mathbf{V}'\in\mathcal{V}$ for $\mathbf{g}(\mathbf{V})\preceq\mathbf{0}$.

\begin{lemma}\label{lm:OCO}
With bounded channel gain $B$, we have
\begin{align}
        &\Vert\nabla_{\mathbf{V}^*} f_t(\mathbf{V})\Vert_F \le D,\quad\forall\mathbf{V}\in\mathcal{V},\quad\forall{t,}\label{eq:D}\\
        &\Vert\mathbf{g}(\mathbf{V})-\mathbf{g}(\mathbf{V}')\Vert\le \beta\Vert\mathbf{V}-\mathbf{V}'\Vert_F,\quad\forall\mathbf{V},\mathbf{V}'\in\mathcal{V},\label{eq:beta}\\
        &\Vert\mathbf{g}(\mathbf{V})\Vert\leq G,\quad\forall\mathbf{V}\in\mathcal{V},\label{eq:G}\\
        &\Vert\mathbf{V}-\mathbf{V}'\Vert_F\leq R,\quad\forall\mathbf{V},\mathbf{V}'\in\mathcal{V},\label{eq:R}\\
        &\exists\mathbf{V}'\in\mathcal{V}, \quad\mathbf{g}(\mathbf{V}')\preceq-\epsilon\mathbf{1}
\end{align}
where $D=B^2R$, $\beta=2\sqrt{\max_{c\in\{1,\dots,C\}}\{P_c^{\text{max}}\}}$,
$G=\sqrt{\sum_{c=1}^C\max\{\bar{P}_c^2,(P_c^{\text{max}}-\bar{P}_c)^2\}}$,
$R=2\sqrt{\sum_{c=1}^CP_c^{\text{max}}}$, and $\epsilon=\min_{c\in\{1,\dots,C\}}\{\bar{P}_c\}$.
\end{lemma}
\textit{Proof:} See Appendix \ref{app:OCO}.

Using Lemma~\ref{lm:OCO}, we provide performance bounds for our online coordinated precoder solution in the following theorem. 
\begin{theorem}\label{thm:BD}
Let $\alpha=\sqrt{\frac{T}{\tau}},\gamma^2=\sqrt{T}$ and $\eta=\frac{1}{2}\beta\sqrt{T}$, then the following statements hold for $\{\tilde{\mathbf{V}}_t^c\}$ in (\ref{eq:Vtc}):
\begin{align}
        &\!\!\!\!\!\text{RE}(T)\!\le\!\frac{D^{2}T}{\alpha}\!+\!\frac{\gamma^2G^2}{2}\!+\!(\alpha\tau\!+\!\eta)R^2\!+\!2DR\tau\!=\!\mathcal{O}(\sqrt{\tau{T}}),\!\!\!\!\label{eq:reg}\\
        &\!\!\!\!\!\text{VO}^c(T)\!\le\!2G\!+\!\frac{2\gamma^2G^2\!+\!2DR\!+\!(\alpha\!+\!\eta)R^2}{\gamma^{2}\epsilon}\!=\!\mathcal{O}(1),~\forall{c}.\label{eq:vio}
\end{align}
\end{theorem}
\textit{Proof:} The proof of (\ref{eq:reg}) is similar to the proof of Theorem~5
in \cite{COCO-J.Wang'21}, except that the problem is in the complex domain, and we need to apply some properties of complex matrix operation. Using Lemma~6 in \cite{COCO-J.Wang'21}, we have $\text{VO}^c(T)\le\frac{1}{\gamma}\Vert\mathbf{Q}_T\Vert+\sum_{t=1}^{\tau}g^c(\tilde{\mathbf{V}}_t^c)$. By initializing $\mathbf{g}(\mathbf{V}_t)=\mathbf{0}$ for $t\le\tau$ and adopting the proof of Theorem~7 in~\cite{COCO-J.Wang'21}, we have~(\ref{eq:vio}).\endIEEEproof

Theorem~\ref{thm:BD} shows that, even with long-term constraints (\ref{eq:LTC}), the $T$-slot regret in (\ref{eq:regdef}) grows in the order of $\mathcal{O}(\sqrt{\tau{T}})$, which is the same as the current best regret for OCO with $\tau$-slot delay subject to the short-term constraints only in \cite{OCO-L.John'09}.  Furthermore, even under $\tau$-slot delay, the $\mathcal{O}(1)$ constraint violation in (\ref{eq:vio}) is the same as the current best constraint violation bound for constrained OCO with one-slot delay in \cite{COCO-H.Yu'20}.

\section{Simulation Results}
\label{Sec:4}

We consider a virtualized MIMO network consisting of $C=3$ urban hexagon micro cells, each with radius $R_c=500$~m. Each BS $c$ is equipped with $N_c=32$ antennas as default. The InP serves $M=4$ SPs, each has $K_c^m=2$ subscribing users randomly located in each cell~$c$. Following typical LTE settings, we focus on a channel with $B_W=15$~kHz bandwidth. We set the maximum transmit power limit $P_c^{\text{max}}=33$~dBm. By default, the time-averaged transmit power limit is $\bar{P}_c=30$~dBm. The receiver thermal noise power spectral density is set to $N_0=-174$~dBm/Hz and the noise figure is $N_F=10$~dB. The fading channel between the $k$-th user of SP $m$ in cell $l$ and the BS $c$ is modeled as a first-order Gauss-Markov process $\mathbf{h}_{t+1}^{lc,m,k}=\alpha_{\mathbf{h}}\mathbf{h}_t^{lc,m,k}+\mathbf{z}_t^{lc,m,k}$, where $\mathbf{h}_t^{lc,m,k}\sim\mathcal{CN}(\mathbf{0},\beta^{lc,m,k}\mathbf{I})$ with $\beta^{lc,m,k}$ representing path-loss and shadowing, $\alpha_{\mathbf{h}}$ is the channel correlation coefficient, and $\mathbf{z}_t^{lc,m,k}\sim\mathcal{CN}(\mathbf{0},(1-\alpha_{\mathbf{h}}^2)\beta^{lc,m,k}\mathbf{I})$ is independent of $\mathbf{h}_t^{lc,m,k}$. We set $\alpha_{\mathbf{h}}=0.998$, which corresponds to pedestrian user speed $1$~km/h.\footnote{For different values of $\alpha_{\mathbf{h}}$, our simulation results are similar and hence are omitted for brevity.} We set the time slot duration to $\frac{1}{B_W}$, and the total time horizon $T=1000$ time slots.

We assume each SP $m$ uses the zero forcing (ZF) precoding scheme to design its virtual precoder $\mathbf{W}_t^{c,m}=\omega_t^{c,m}\mathbf{H}_t^{cc,m}(\mathbf{H}_t^{cc,m}{\hbox{$\mathbf{H}_t^{cc,m}$}}^H)^{-1}$, where $\omega_t^{c,m}$ is a power scaling factor such that $\Vert\mathbf{W}_t^{c,m}\Vert_F^2=P_c^m=\frac{P_c^{\text{max}}}{M}$. For performance evaluation, we define the time-averaged normalized precoding deviation as $\bar{f}(T)\triangleq\frac{1}{T}\sum_{t=1}^T\frac{f_t(\mathbf{V}_t)}{\Vert\mathbf{D}_t\Vert_F^2}$, the time-averaged transmit power as $\bar{P}(T)\triangleq\frac{1}{TC}\sum_{t=1}^T\Vert\mathbf{V}_t\Vert_F^2$, and the time-averaged per-user rate as $\bar{R}(T)\triangleq\frac{1}{TK}\sum_{t=1}^T\sum_{k=1}^K\log_2(1+\frac{|[\mathbf{H}_t\mathbf{V}_t]_{k,k}|^2}{\sum_{i=1,i\neq{k}}^K|[\mathbf{H}_t\mathbf{V}_t]_{k,i}|^2+\sigma_n^2})$, where $\sigma_n^2=N_0B_W+N_F$, and $[\mathbf{A}]_{i,j}$ denotes the $(i,j)$ element of matrix $\mathbf{A}$. For performance comparison, we consider 1) the online algorithm in \cite{COCO-X.CAO'20}, which is currently the best constrained OCO algorithm that accommodates multi-slot feedback delay; 2) the standard offline benchmark $\mathbf{V}^\star$; 3) a frequency division (FD) scheme that allocates equal bandwidth $\frac{B_W}{M}$ to each SP $m$. This FD approach is commonly adopted in the existing literature of MIMO WNV \cite{WNV-V.Jumba'15}\nocite{WNV'K.Zhu'16}-\cite{WNV-Y.Liu'18}. At each cell $c$, each SP $m$ adopts ZF precoding to serve its $K_c^m$ users directly based on the local~CSI.

\begin{figure}[t]
\vspace{-2mm}
\centering
\includegraphics[width=.49\linewidth,trim= 00 00 00 00,clip]{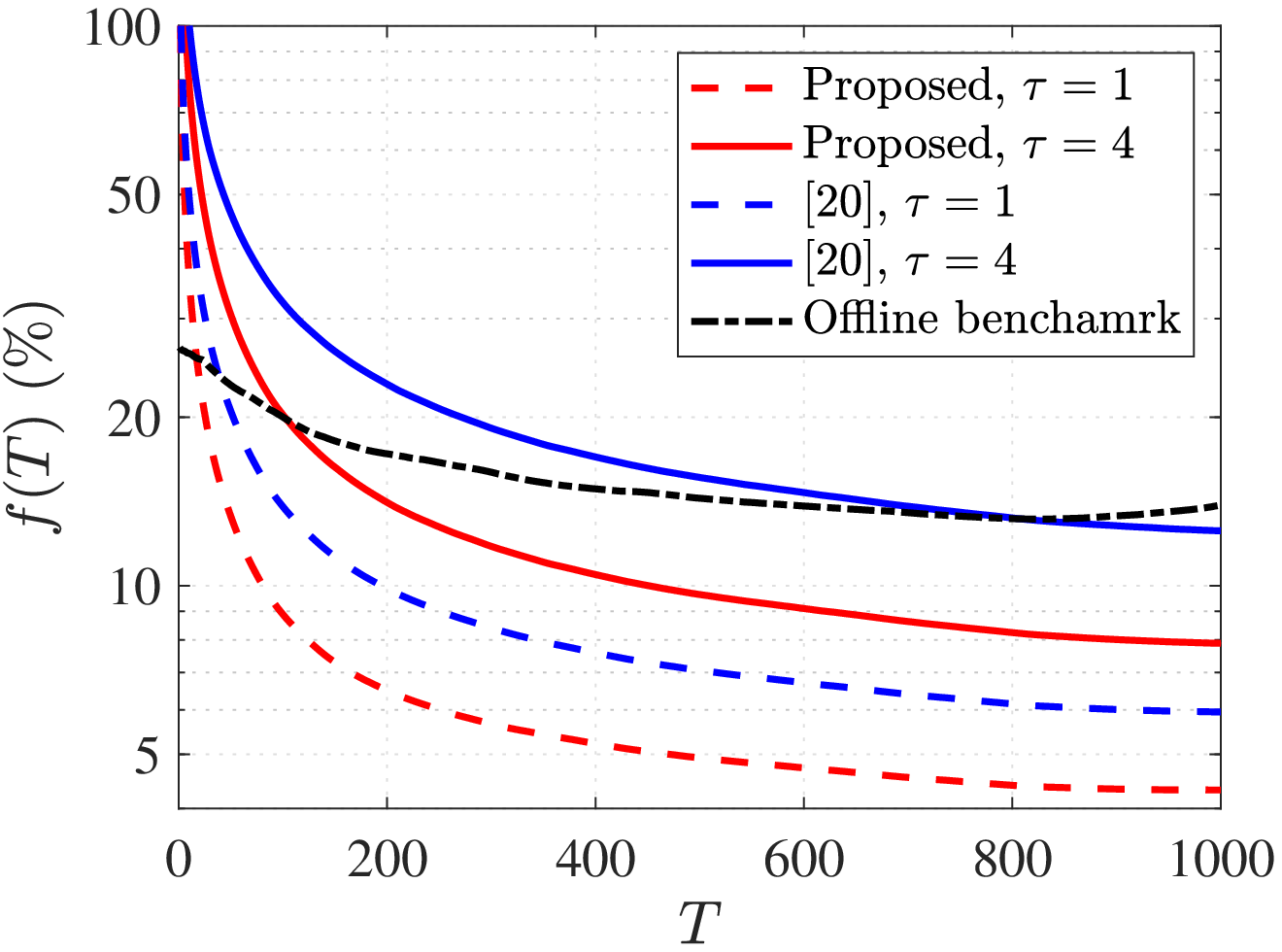}
\includegraphics[width=.49\linewidth,trim= 00 00 00 00,clip]{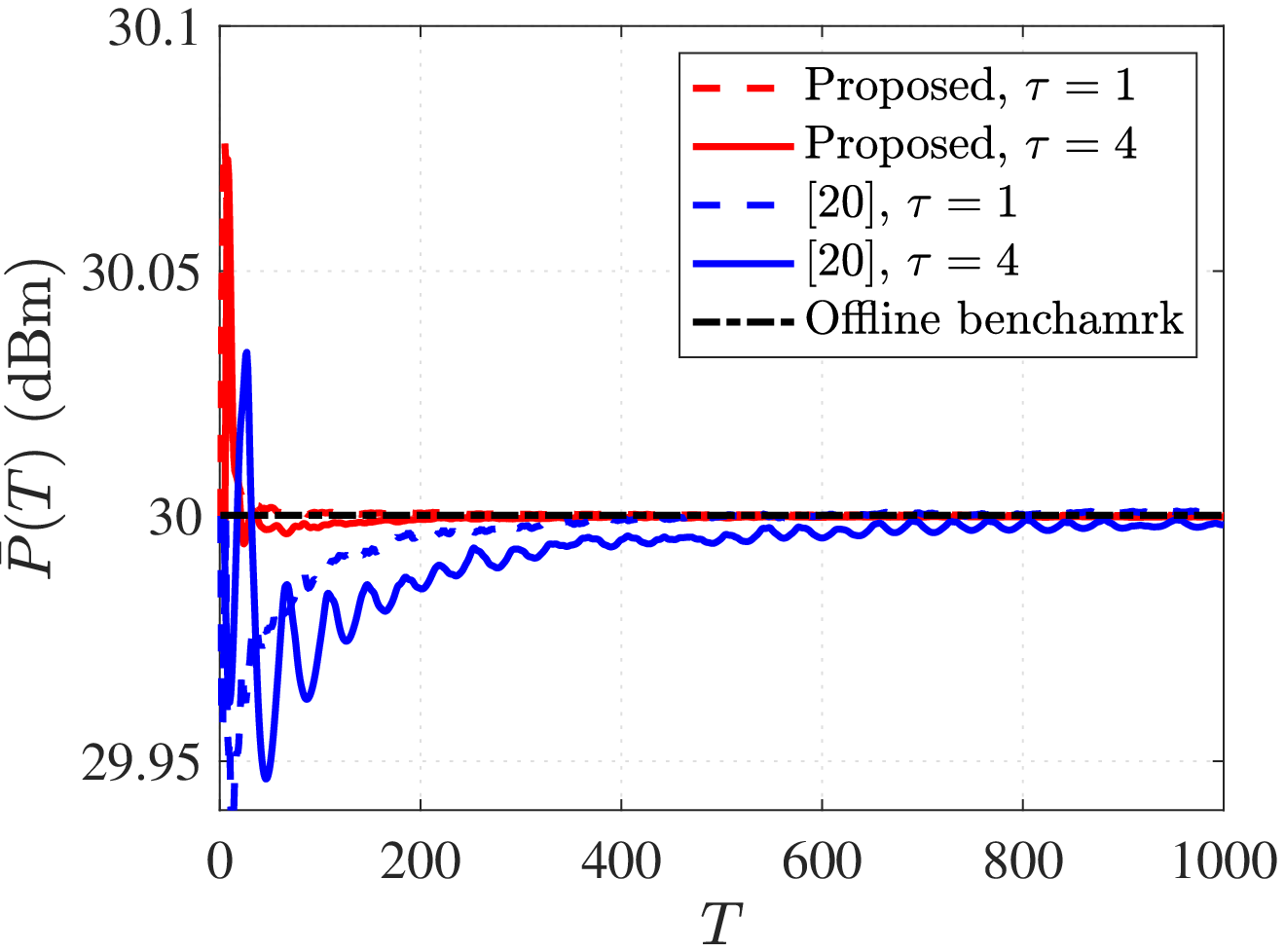}
\vspace{-5mm}
\caption{$\bar{f}(T)$ and $\bar{R}(T)$ vs. $T$ under different
$\tau$ values.}
\label{fig:2}
\vspace{-3mm}
\end{figure}

Fig.~\ref{fig:2} shows $\bar{f}(T)$ and $\bar{P}(T)$ versus $T$ with different values of CSI feedback delay $\tau$. Our online precoding solution outperforms the one in \cite{COCO-X.CAO'20}, which uses the gradient $\nabla\mathbf{g}(\mathbf{V})$ information to minimize $\mathbf{g}(\mathbf{V})$. In our solution, we directly minimize the constraint function $\mathbf{g}(\mathbf{V})$ as one part of the objective function in \textbf{P2}, which improves the control of the transmit power. Furthermore, the regularization is imposed on both $\mathbf{V}_{t-\tau}$ and $\mathbf{V}_{t-1}$ instead of the single regularization term in \cite{COCO-X.CAO'20}. The double regularization improves the efficiency of gradient descent on minimizing the precoding deviation.

\begin{figure}[!t]
\vspace{-2mm}
\centering
\includegraphics[width=.49\linewidth,trim= 00 00 00 00,clip]{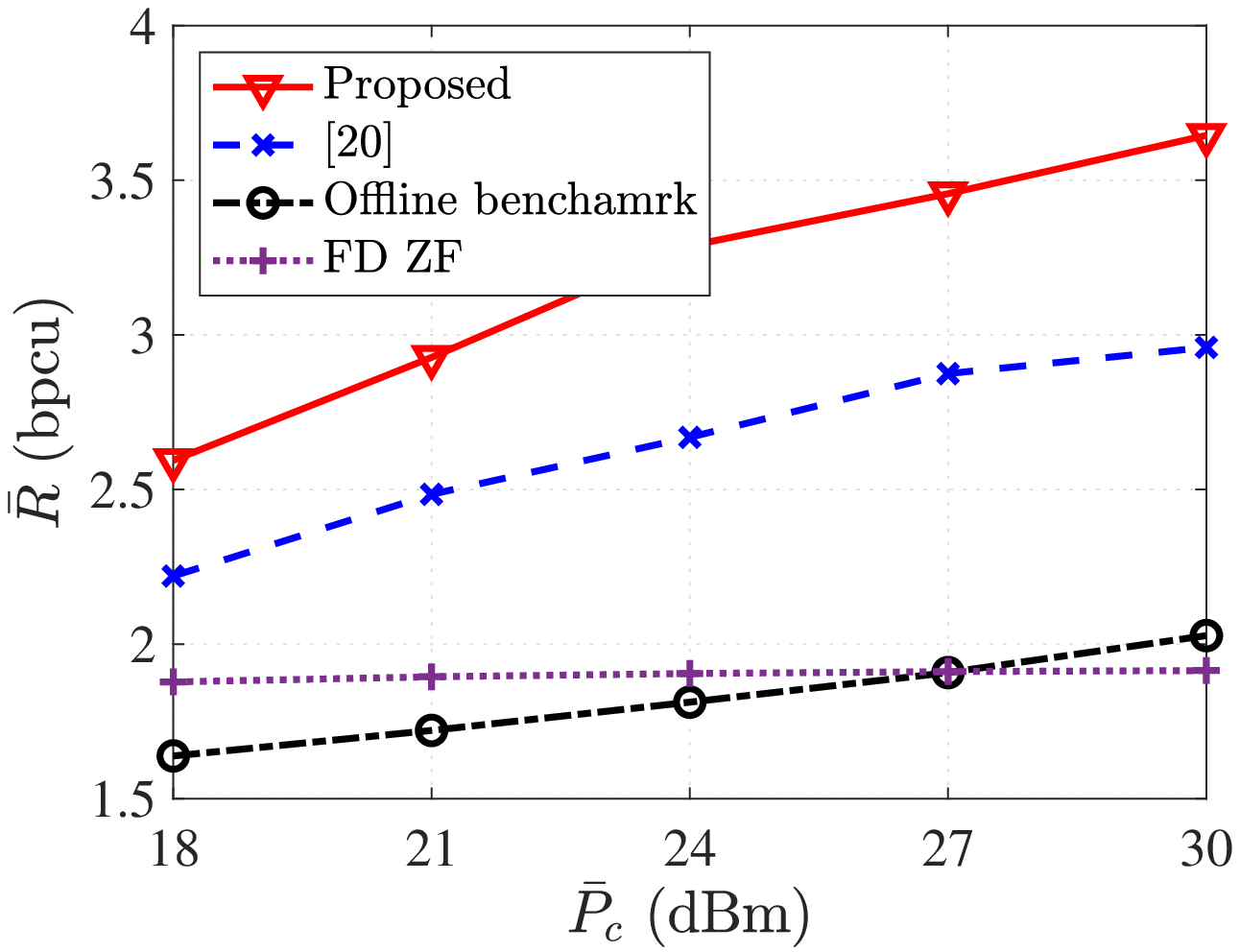}
\includegraphics[width=.49\linewidth,trim= 00 00 00 00,clip]{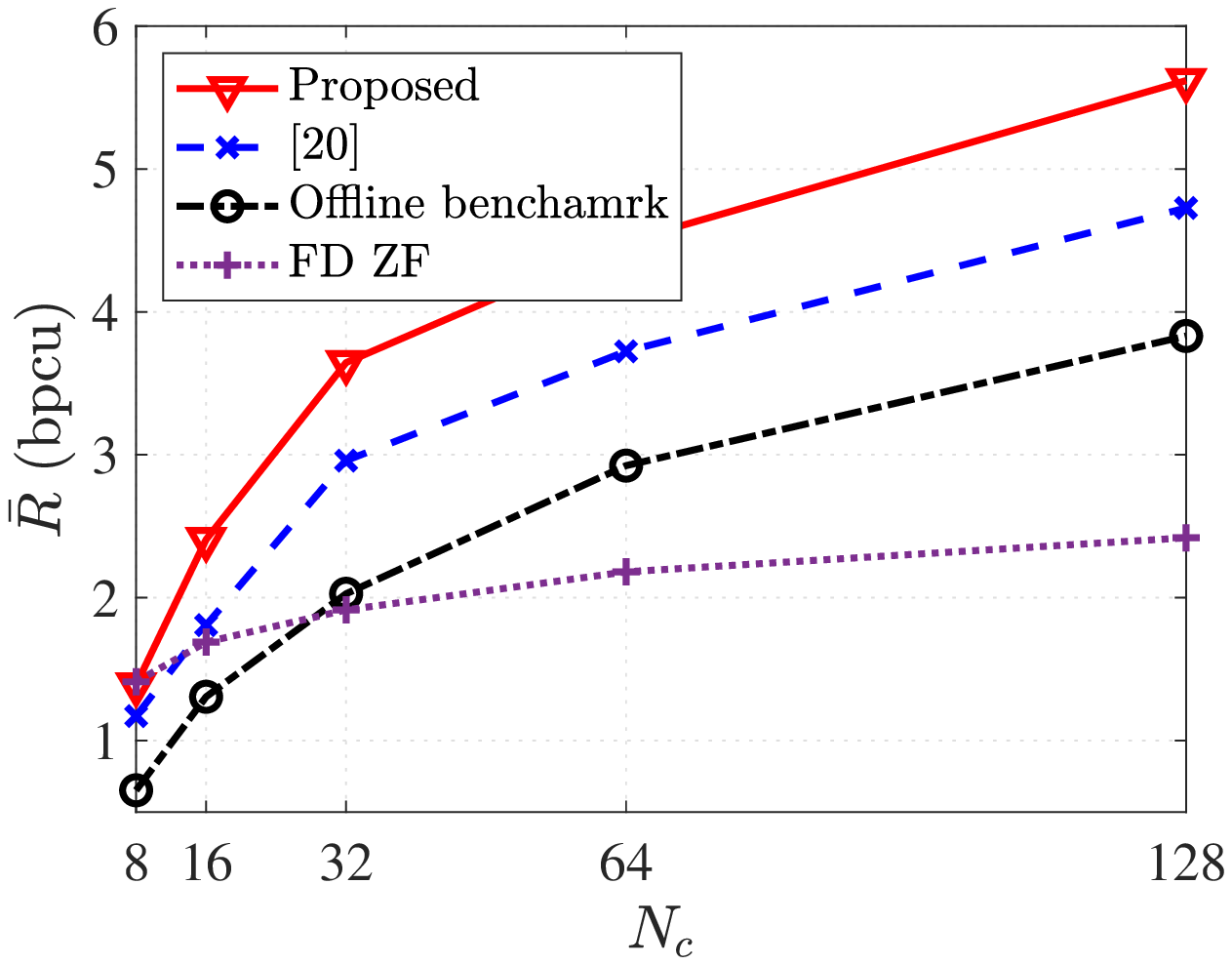}
\vspace{-5mm}
\caption{Impacts of $\bar{P}_c$ and $N_c$ on $\bar{R}$ with $\tau=4$.}
\label{fig:3}
\vspace{-3mm}
\end{figure}

We further study the impacts of the long-term transmit power limit $\bar{P}^c$ and $N_c$ on the performance of our precoding solution. We set $\tau=4$. Fig.~\ref{fig:3} shows that the steady-state per-user rate $\bar{R}$ increases as $\bar{P}^c$ increases. This is because the InP can use more transmit power to mitigate interference and meet the virtualization demand. \blue{} Furthermore, $\bar{R}$ increases as $N_c$ increases, indicating the effectiveness of the precoding solution in interference mitigation. When $N_c$ is large, our proposed online solution substantially outperforms the FD ZF scheme. This demonstrates the effectiveness of the proposed spatial isolation approach via simultaneously sharing all frequency resources among SPs. We observe that in a wide range of $\bar{P}_c$ and $N_c$ values, our proposed precoding solution substantially outperforms the online solution from \cite{COCO-X.CAO'20}, the offline benchmark, and the FD ZF scheme.

\section{Conclusions}

In this letter, we have considered online coordinated precoding design for multi-cell MIMO WNV with delayed CSI. Our goal is to minimize the accumulated deviation of the InP's precoder from the SPs' virtualization demands over time, subject to both long-term and short-term per-cell transmit power constraints. We have developed an online coordinated precoding solution with provable performance bounds. Our precoding solution is in closed-form and is fully distributed at each cell. Simulation reveals substantial performance gain of our precoding solution over the current best alternative. 

\appendices
\section{Proof of Lemma~\ref{lm:OCO}}\label{app:OCO}

\textit{Proof}: We first show (\ref{eq:D}). We have
\begin{align}
        &\Vert\nabla_{\mathbf{V}^*}f_t(\mathbf{V})\Vert_F=\Vert\mathbf{H}_t^H(\mathbf{H}_t\mathbf{V}-\mathbf{D}_t)\Vert_F\notag\\
        &\stackrel{(a)}{\le}\Vert\mathbf{H}_t\Vert_F\Vert\mathbf{H}_t\mathbf{V}-\mathbf{D}_t\Vert_F\stackrel{(b)}{\le}\Vert\mathbf{H}_t\Vert_F(\Vert\mathbf{H}_t\mathbf{V}\Vert_F+\Vert\mathbf{D}_t\Vert_F)\notag\\
        &\le\Vert\mathbf{H}_t\Vert_F(\Vert\mathbf{H}_t\Vert_F\Vert\mathbf{V}\Vert_F+\Vert\mathbf{D}_t\Vert_F)\stackrel{(c)}{\le}2B^{2}\sqrt{\sum_{c=1}^CP_{\text{max}}^c}
\end{align}
where $(a)$ follows from inequality $\Vert\mathbf{A}\mathbf{B}\Vert_F\le\Vert\mathbf{A}\Vert_F\Vert\mathbf{B}\Vert_F$,
$(b)$ is because $\Vert\mathbf{A}+\mathbf{B}\Vert_F\le\Vert\mathbf{A}\Vert_F+\Vert\mathbf{B}\Vert_F$,
and $(c)$ follows from $\Vert\mathbf{H}_t\Vert_F\le{B},\forall{t}\in\{1,\dots,T\}$
by assumption, $\mathbf{V}\in\mathcal{V}\triangleq\{\mathbf{V}:\Vert\tilde{\mathbf{V}}^c\Vert_F^2\le{P}_c^\text{max},\forall{c}\in\{1,\dots,C\}\}$
such that $\Vert\mathbf{V}\Vert_F^2\le\sum_{c=1}^C\Vert\tilde{\mathbf{V}}^c\Vert_F^2\le\sum_{c=1}^CP_c^{\text{max}}$,
and
\begin{align*}
        \Vert\mathbf{D}_t\Vert_F^2&=\sum_{c=1}^C\Vert\mathbf{D}_t^c\Vert_F^2=\sum_{c=1}^C\sum_{m=1}^M\Vert\mathbf{H}_t^{cc,m}\mathbf{W}_t^{c,m}\Vert_F^2\notag\\
        &\leq\sum_{c=1}^C\sum_{m=1}^M\Vert\mathbf{H}_t^{cc,m}\Vert_F^2\Vert\mathbf{W}_t^{c,m}\Vert_F^2\le{B^{2}}\sum_{c=1}^CP_{\text{max}}^c.
\end{align*}

From the first order condition of real valued scalar function $g^c(\tilde{\mathbf{V}}^c)$
with respect to complex valued matrix variable $\tilde{\mathbf{V}}^c$, for any
$\tilde{\mathbf{V}}_1^c$ and $\tilde{\mathbf{V}}_2^c$, we have
\begin{align*}
        &g^c(\tilde{\mathbf{V}}_1^c)-g^c(\tilde{\mathbf{V}}_2^c)\le2|\tr\{[\nabla_{\tilde{\mathbf{V}}_1^{c*}}g^c(\tilde{\mathbf{V}}_1^c)]^H(\tilde{\mathbf{V}}_2^c-\tilde{\mathbf{V}}_1^c)\}|\notag\\
        &\le-2\Re\{\tr\{[\nabla_{\tilde{\mathbf{V}}_1^{c*}}g^c(\tilde{\mathbf{V}}_1^c)]^H(\tilde{\mathbf{V}}_2^c-\tilde{\mathbf{V}}_1^c)\}\}\notag\\
        &\le2\Vert\nabla_{\tilde{\mathbf{V}}_1^{c*}}g^c(\tilde{\mathbf{V}}_1^c)\Vert_F\Vert\tilde{\mathbf{V}}_2^c-\tilde{\mathbf{V}}_1^c\Vert_F\le2\Vert\tilde{\mathbf{V}}_1^c\Vert_F\Vert\tilde{\mathbf{V}}_2^c-\tilde{\mathbf{V}}_1^c\Vert_F\notag\\
        &\le2\sqrt{P_{\text{max}}^c}\Vert\tilde{\mathbf{V}}_2^c-\tilde{\mathbf{V}}_1^c\Vert_F.
\end{align*}
Taking square on both sides of the above inequality and summing over $c\in\{1,\dots,C\}$,
we have
\begin{align}
        \!\Vert\mathbf{g}(\mathbf{V}_1)-\mathbf{g}(\mathbf{V}_2)\Vert_2^2&\le\sum_{c=1}^C4P_{\text{max}}^c\Vert\tilde{\mathbf{V}}_2^c-\tilde{\mathbf{V}}_1^c\Vert_F^2\notag\\
        \!&\le4\max_{c\in\{1,\dots,C\}}\{P_{\text{max}}^c\}\Vert\mathbf{V}_1-\mathbf{V}_2\Vert_F^{2},\!
\end{align}
which yields (\ref{eq:beta}).
 
From $\Vert\tilde{\mathbf{V}}^c\Vert_F^2\le{P}_{\text{max}}^c$ and $g^c(\tilde{\mathbf{V}}^c)\triangleq\Vert\tilde{\mathbf{V}}^c\Vert_F^2-\bar{P}^c,\forall{c}\in\{1,\dots,C\}$,
we have
\begin{align}
        \!\!\!\Vert\mathbf{g}(\mathbf{V})\Vert_2^2=\!\!\sum_{c=1}^C[g^c(\tilde{\mathbf{V}}^c)]^2\!\leq\!\sum_{c=1}^C\!\max\{\bar{P}^{c2},(P_{\text{max}}^c\!-\!\bar{P}^c)^2)\},\!\!
\end{align}
which gives (\ref{eq:G}).

We now show (\ref{eq:R}). For any $\mathbf{V}_1,\mathbf{V}_2\in\mathcal{V}$, we have
\begin{align}
        \Vert\mathbf{V}_1-\mathbf{V}_2\Vert_F\leq\Vert\mathbf{V}_1\Vert_F+\Vert\mathbf{V}_2\Vert_F\leq2\sqrt{\sum_{c=1}^CP_{\text{max}}^c}.
\end{align}

Finally, let $\mathbf{V}'=\mathbf{0}$, we have
\begin{align}
        \mathbf{g}(\mathbf{V}')\preceq-\mathbf{1}\cdot\min_{c\in\{1,\dots,{C}\}}\{\bar{P}^c\}
\end{align}
where $\mathbf{1}$ is a vector
of all $1$'s.\endIEEEproof

\bibliographystyle{IEEEtran}
\bibliography{Reference}

\end{document}